\documentstyle[twocolumn,twoside]{article}
\newcommand{\AAA}{{\cal A \hspace*{-1.7ex} A}}
\newcommand{\BB}{{\cal B \hspace*{-1.7ex} B}}
\newcommand{\DD}{{\cal D \hspace*{-2ex} D}}
\newcommand{\dotnabla}{\nabla \hspace*{-1.26ex} \raisebox{0.4ex}{$\cdot \,$}}

\newcommand{\FF}{{\cal F \hspace*{-2ex} F}}
\newcommand{\II}{{\cal I \hspace*{-1.3ex} I}}

\newcommand{\KK}{{\cal K \hspace*{-1.6ex} K}}
\newcommand{\MM}{{\cal M \hspace*{-2.7ex} M}}
\newcommand{\RR}{{\cal R \hspace*{-2.1ex} R}}
\newcommand{\SS}{{\cal S \hspace*{-1.7ex} S}}
\newcommand{\TT}{{\cal T \hspace*{-1.7ex} T}}
\newcommand{\WW}{{\cal W \hspace*{-2.6ex} W}}
\title{\Large \bf Five-Dimensional Tangent Vectors in Space-Time \\
\large \bf III. Some Applications}
\author{Alexander Krasulin \\ \it Institute for Nuclear Research of the
Russian Academy of Sciences \\ \it 60th October Anniversary Prospect, 7a,
117312 Moscow, Russia}
\date{\normalsize \bf Abstract \\ \mbox{ } \\ \begin{minipage}{400pt}
\normalsize In this part of the series I show how five-tensors can be used
for describing in a coordinate-independent way finite and infinitesimal
Poincare transformations in flat space-time. As an illustration, I
reformulate the classical mechanics of a perfectly rigit body in terms of
the analogs of five-vectors in three-dimensional Euclidean space. I then
introduce the notion of the bivector derivative for scalar, four-vector
and four-tensor fields in flat space-time and calculate its analog in
three-dimensional space for the Lagrange function of a system of several
point particles in classical nonrelativistic mechanics. \end{minipage} }
\begin{document}

\maketitle

\begin{flushleft}
A. \it Preliminary remarks
\end{flushleft}
Before developing the mathematical theory of five-vectors further, it
will be useful to consider some of their applications in flat space-time.
At the same time I will say a few words about the analogs of five-vectors
in three-dimensional Euclidean space and will consider some of their
applications.

In the case of vectors of the latter type there arises a problem with
terminology. Since the analogs of five-vectors in three-dimensional space
are four-dimensional, by analogy with five-vectors one should term them as
{\em four}-vectors. This, of course, is unacceptable since the term
`four-vector' is traditionally used for referring to ordinary tangent
vectors in space-time. To avoid confusion, in the following the analogs of
five-vectors in three-dimensional space will be called (3+1)-vectors. It
makes sense to use the same terminology for all other manifolds as well.
The only exception from this convention will be made for five-vectors in
space-time, which will still be called five-vectors and not (4+1)-vectors.

Let me also say a few words about the notations. Ordinary tangent vectors in
three-dimensional Euclidean space will be denoted with capital Roman letters
with an arrow: $\vec{A}$, $\vec{B}$, $\vec{C}$, etc. Three-plus-one-vectors
will be denoted with lower-case Roman letters with an arrow: $\vec{a}$,
$\vec{b}$, $\vec{c}$, etc. It will be taken that lower-case latin indices
run 1, 2, and 3 and that capital Greek indices run 1, 2, 3, and 5 (the value
``5'' corresponds to the additional dimension of the space of (3+1)-vectors).
Finally, the Euclidean inner product will be denoted with a dot or with
the symbol $\delta$, which should not be confused with the unit tensor of
rank $(1,1)$.

In conclusion, let me write out the formulae that express the result of
transporting parallelly an active regular five-vector basis associated with
some system of Lorentz coordinates, from the origin of the latter to the
point with coordinates $x^{\mu}$:
\begin{equation} \left\{ \begin{array}{l}
{[}{\bf e}_{\alpha}(0){]}^{{\rm transported \; to}\; x} =
{\bf e}_{\alpha}(x) + x_{\alpha} \, {\bf e}_{5}(x) \\
{[}{\bf e}_{5}(0){]}^{{\rm transported \; to}\; x}  = {\bf e}_{5}(x).
\end{array} \right. \end{equation}
The analogs of these formulae for (3+1)-vectors are
\begin{equation} \left\{ \begin{array}{l}
{[}\vec{e}_{i}(0){]}^{{\rm transported \; to}\; x} = \vec{e}_{i}(x) + x_{i}
\, \vec{e}_{5}(x) \\ {[} \vec{e}_{5} (0) {]}^{{\rm transported \; to}\; x}
= \vec{e}_{5}(x),
\end{array} \right. \end{equation}
where $x_{i} \equiv \delta_{ij} x^{j}$ and $x^{j}$ is the system of Cartesian
coordinates with which the considered active regular basis of (3+1)-vectors
is associated.

\vspace{3ex} \begin{flushleft}
B. \it Active Poincare transformations \\
\hspace*{2.5ex} in the language of five-vectors
\end{flushleft}
The replacement of any given set of scalar, four-vector and four-tensor
fields in flat space-time with an equivalent set of fields, as it was
discussed in section 4 of part I, and a similar replacement of
five-vector and five-tensor fields, are nothing but active Poincare
transformations. Though the procedure for constructing the equivalent fields,
described in the same section of part I for four-vector and four-tensor
fields and in section 3 of part II for five-vector and five-tensor
fields, makes use of Lorentz coordinate systems, the latter are only a
tool for performing such a construction. In itself, any such active field
transformation can be considered without referring to any coordinates and
in this sense is an invariant operation. Let us now see how one can give
it a coordinate-independent description.

 From the very procedure for constructing the equivalent fields it follows
that any active Poincare transformation can be characterised by the
parameters $L^{\alpha}_{\; \beta}$ and $a^{\alpha}$ that specify the
transition from the selected initial Lorentz coordinate system $x^{\alpha}$
to the corresponding final Lorentz coordinate system $x'^{\, \alpha}$:
\begin{equation}
x'^{\, \alpha} = L^{\alpha}_{\; \beta} \, x^{\beta} + a^{\alpha}.
\end{equation}
Since parameters $L^{\alpha}_{\; \beta}$ and $a^{\alpha}$ explicitly depend
on the choice of the initial coordinate system, such a description will
naturally be non-invariant. To reduce the dependence on the choice of the
coordinate system, instead of $L^{\alpha}_{\; \beta}$ and $a^{\alpha}$ one
speaks of the displacement vector and rotation tensor. Namely, in the
selected initial system of Lorentz coordinates, from the corresponding
transformation parameters $L^{\alpha}_{\; \beta}$ and $a^{\alpha}$,
basis four-vectors ${\bf E}_{\alpha}$, and dual basis four-vector
1-forms $\widetilde{\bf O}^{\alpha}$ one constructs the four-vector
${\bf A} \equiv a^{\alpha} {\bf E}_{\alpha}$ and four-tensor ${\bf B} \equiv
( L^{\alpha}_{\; \beta} - \delta^{\alpha}_{\, \beta} ) \, {\bf E}_{\alpha}
\otimes \widetilde{\bf O}^{\alpha}$. By using the transformation formulae
for $L^{\alpha}_{\; \beta}$ and $a^{\alpha}$ derived in section 5 of part I,
one can easily show that $\bf B$ will be the same at any choice of the
initial Lorentz coordinate system and that $\bf A$ will be the same for
any two systems with the same origin. For systems with different origins
$\bf A$ will in general be different, so by using the four-vector quantities
$\bf A$ and $\bf B$ one cannot get rid of the dependence on the choice of
the initial Lorentz coordinate system completely.

To obtain a completely coordinate-independent description of a given active
Poincare transformation, let us construct in the selected Lorentz coordinate
system a five-tensor field $\TT$ which in the $P$-basis associated with these
coordinates has the following constant components:
\begin{equation} \left\{ \begin{array}{ll}
{\cal T}^{\alpha}_{\; \beta} = \Lambda^{\alpha}_{\; \beta}, \; &
{\cal T}^{\alpha}_{\; 5} = 0 \\ {\cal T}^{5}_{\; \beta} = a_{\beta}, &
{\cal T}^{5}_{\; 5} = 1,
\end{array} \right. \end{equation}
where $a_{\beta} \equiv g_{\beta \sigma} a^{\sigma}$ and
$\Lambda^{\alpha}_{\; \beta} \equiv (L^{-1})^{\alpha}_{\; \beta}$ is the
inverse of the $4 \times 4$ matrix $L^{\alpha}_{\; \beta}$. As it has been
shown in section 5 of part I, the quantities ${\cal T}^{A}_{\; B}$
transform as components of a five-tensor of rank $(1,1)$, and consequently
the field $\TT$ will be the same at any choice of the initial Lorentz
coordinate system.

To understand why the components of $\TT$ are constructed from the
transformation parameters for covariant coordinates and not from the
transformation parameters for the Lorentz coordinates themselves, let us
introduce this five-tensor in a slightly different way. Instead of the two
coordinate systems $x^{\alpha}$ and $x'^{\, \alpha}$, let us consider the
corresponding $P$-bases, ${\bf p}_{A}$ and ${\bf p}'_{A}$. Since at each
point the vectors ${\bf p}'_{A}$ are expressed linearly in terms of
${\bf p}_{A}$, the former can be obtained by acting on the latter with some
linear operator, $\TT$. Since all $P$-bases are self-parallel by definition,
$\TT$ regraded as a five-tensor field of rank $(1,1)$ will be covariantly
constant. If $\widetilde{\bf q}^{A}$ and $\widetilde{\bf q}'^{\, A}$ are
the bases of five-vector 1-forms dual to ${\bf p}_{A}$ and ${\bf p}'_{A}$,
respectively, one can presented the tensor $\TT$ as
\begin{displaymath}
{\TT} = {\bf p}'_{A} \otimes \widetilde{\bf q}^{A},
\end{displaymath}
for in this case
\begin{displaymath}
{\TT}({\bf p}_{A}) \equiv {\bf p}'_{B}
< \widetilde{\bf q}^{B}, {\bf p}_{A} > \; = {\bf p}'_{A}.
\end{displaymath}
Furthermore,
\begin{displaymath}
{\TT} ( \widetilde{\bf q}'^{\, A}) = \; < \widetilde{\bf q}'^{\, A},
 {\bf p}'_{B} > \widetilde{\bf q}^{B} = \widetilde{\bf q}^{A},
\end{displaymath}
so when acting on 1-forms, the operator $\TT$ performs the reverse
transformation. The inverse of $\TT$ is apparently ${\TT}^{-1} = {\bf p}_{A}
\otimes \widetilde{\bf q}'^{\, A}$.

It is not difficult to prove that the field $\TT$ defined this way will be
the same at any choice of the initial $P$-basis. Indeed, when acting on the
fields ${\bf p}_{A}$ of the initial $P$-basis, the operator $\TT$ performs
the considered active field transformation, and since this transformation
is linear, $\TT$ will act the same way on all other covariantly constant
five-vector fields, including the basis fields of any other $P$-basis.

Let us now find the components of $\TT$ and ${\TT}^{-1}$ in the basis
${\bf p}_{A} \otimes \widetilde{\bf q}^{B}$. Since
\begin{displaymath}
{\bf p}'_{\alpha} = {\bf p}_{\beta} \Lambda^{\beta}_{\; \alpha} +
a_{\alpha} {\bf p}_{5} \; \; \mbox{ and } \; \; {\bf p}'_{5} = {\bf p}_{5},
\end{displaymath}
one has
\begin{displaymath} \left. \begin{array}{rcl}
{\TT} & = & ({\bf p}_{\beta} \Lambda^{\beta}_{\; \alpha} + a_{\alpha}
{\bf p}_{5}) \otimes \widetilde{\bf q}^{\alpha} + {\bf p}_{5} \otimes
\widetilde{\bf q}^{5} \\ & = & {\bf p}_{\alpha} \otimes \widetilde{\bf q}^
{\beta} \cdot \Lambda^{\alpha}_{\; \beta} + {\bf p}_{5} \otimes \widetilde
{\bf q}^{\beta} \cdot a_{\beta} \\ & & \hspace{0ex} + \; {\bf p}_{\alpha}
\otimes \widetilde{\bf q}^{5} \cdot 0 + {\bf p}_{5} \otimes
\widetilde{\bf q}^{5} \cdot 1,
\end{array} \right. \end{displaymath}
so in this basis the compoents of $\TT$ are given by formulae (4). In a
similar manner one can find the components of ${\TT}^{-1}$:
\begin{displaymath} \left\{ \begin{array}{ll}
({\cal T}^{-1})^{\alpha}_{\; \beta} = L^{\alpha}_{\; \beta}, \; &
({\cal T}^{-1})^{\alpha}_{\; 5} = 0 \\ ({\cal T}^{-1})^{5}_{\; \beta}
= - \, a_{\gamma} L^{\gamma}_{\; \beta}, & ({\cal T}^{-1})^{5}_{\; 5} = 1.
\end{array} \right. \end{displaymath}
One should not mix the latter up with the quantities
\begin{equation} \left\{ \begin{array}{ll}
{\cal T}^{\alpha}_{\; \beta} = L^{\alpha}_{\; \beta}, \; &
{\cal T}^{\alpha}_{\; 5} = a^{\beta} \\ {\cal T}^{5}_{\; \beta} = 0, &
{\cal T}^{5}_{\; 5} = 1
\end{array} \right. \end{equation}
constructed from the transformation parameters for the Lorentz coordinates
themselves. What the latter are is explained in Appendix.

\vspace{3ex}

Let us now consider an infinitesimal Poincare transformation. In this case
the matrix $L^{\alpha}_{\; \beta}$ in formula (3) can be presented as
\begin{displaymath}
L^{\alpha}_{\; \beta}=\delta^{\alpha}_{\, \beta}+\omega^{\alpha}_{\; \beta},
\end{displaymath}
where $\omega^{\alpha}_{\; \beta}$ are infinitesimals that satisfy the
condition
\begin{equation}
\omega_{\alpha \beta} + \omega_{\beta \alpha} = 0,
\end{equation}
where $\omega_{\alpha \beta} \equiv g_{\alpha \sigma}
\omega^{\sigma}_{\; \beta}$. The parameters $a^{\alpha}$ in formula (3)
are also infinitesimals and accordingly the field $\TT$ that describes this
infinitesimal Poincare transformation can be presented as
\begin{displaymath}
\TT = {\bf 1} - \SS,
\end{displaymath}
where $\bf 1$ is the unity five-tensor and tensor $\SS$ has the components
\begin{equation}
{\cal S}^{\alpha}_{\; \beta} = \omega^{\alpha}_{\; \beta}, \;
{\cal S}^{5}_{\; \beta} = - \, a_{\beta}, \; {\cal S}^{A}_{\; 5} = 0,
\end{equation}
and therefore
\begin{equation}
{\SS} = ( {\bf p}_{\alpha} \omega^{\alpha}_{\; \beta} -
{\bf p}_{5} a_{\beta} ) \otimes \widetilde{\bf q}^{\beta}.
\end{equation}
It is evident that the latter expression for $\SS$ will be valid in any
$P$-basis.

In view of condition (6), it is more convenient to deal not with $\SS$
itself, but with the corresponding five-tensor whose indices are either both
lower or both upper. It is essential that this latter five-tensor should be
related to $\SS$ by the map $\vartheta_{g}$ and not by $\vartheta_{h}$, for
the quantities $\omega_{\alpha \beta}$ in the left-hand side of equation
(6) are obtained from parameters $\omega^{\alpha}_{\; \beta}$ by
contraction with the matrix $g_{\alpha \beta}$, not $h_{AB}$. (Besides,
if one lowers or raises an index of $\SS$ with $h_{AB}$, one will obtain
a covariantly {\em non}constant tensor field.) Since
\begin{equation}
\vartheta_{g}({\bf p}_{\alpha}) = \widetilde{\bf q}^{\beta} \eta_{\beta
\alpha} \; \mbox{ and } \; \vartheta_{g}({\bf p}_{5}) = {\bf 0},
\end{equation}
the completely covariant tensor related to $\SS$ by $\vartheta_{g}$ is
\begin{equation}
\vartheta_{g}({\bf p}_{\alpha} \omega^{\alpha}_{\; \beta} - {\bf p}_{5}
a_{\beta}) \otimes \widetilde{\bf q}^{\beta} = \omega_{\alpha \beta} \;
\widetilde{\bf q}^{\alpha} \otimes \widetilde{\bf q}^{\beta},
\end{equation}
and is apparently the five-vector equivalent of the completely covariant
four-tensor of infinitesimal rotation: ${\bf R} = \omega_{\alpha \beta} \;
\widetilde{\bf O}^{\alpha} \otimes \widetilde{\bf O}^{\beta}$. However,
tensor (10) has only six independent components (which is a consequence of
$\vartheta_{g}$ being noninjective) and therefore does not describe the
considered infinitesimal Poincare transformation completely.

Let us now observe that according to formulae (8) and (9), $\SS$ can be
obtained from a completely contravariant five-tensor of rank 2 by lowering
one of the indices of the latter with $g_{\alpha \beta}$. Indeed, denoting
this contravariant five-tensor as $\RR$, by virtue of formulae (9) one has
\begin{displaymath}
{\cal R}^{AB} \, {\bf p}_{A} \otimes \vartheta_{g}({\bf p}_{B}) =
( {\bf p}_{\alpha} {\cal R}^{\alpha}_{\; \beta} + {\bf p}_{5}
{\cal R}^{5}_{\; \beta} ) \otimes \widetilde{\bf q}^{\beta},
\end{displaymath}
where ${\cal R}^{A}_{\; \beta} \equiv {\cal R}^{A \xi} g_{\xi \beta}$,
and comparing the right-hand side of the latter equation with formula (8)
one finds that
\begin{equation}
{\cal R}^{\alpha \beta} = \omega^{\alpha}_{\; \xi} \, g^{\, \xi \beta} \equiv
\omega^{\alpha \beta} \; \mbox{ and } \; {\cal R}^{5 \beta} = - \, a^{\beta}.
\end{equation}
The components ${\cal R}^{A5}$ are not fixed by the components of $\SS$ and
can be selected arbitrarily. A particularly convenient choice is
\begin{equation}
{\cal R}^{\alpha 5} = - \, {\cal R}^{5 \alpha} = a^{\alpha}
\; \mbox{ and } \; {\cal R}^{55} = 0.
\end{equation}
In this case $\RR$ becomes completely antisymmetric and consequently has only
10 independent components, i.e.\ exactly as many as does the five-tensor
$\SS$. Since antisymmetry is an invariant property, equations (12) will
hold in any five-vector basis. Moreover, $\RR$ will be a covariantly constant
field. By analogy with the formula for generators of Lorentz transformations
for four-vectors, the relation between $\RR$ and $\SS$ can be presented as
\begin{equation}
{\cal S}^{A}_{\, B} = - \, {\scriptstyle \frac{1}{2}} \, {\cal R}^{KL} \,
(M_{KL})^{A}_{\, B} = - {\cal R}^{|KL|} \, (M_{KL})^{A}_{\, B},
\end{equation}
where, as usual, the vertical bars around the indices mean that summation
extends only over $K < L$, and the quantities
\begin{equation}
(M_{KL})^{A}_{\, B} \equiv \delta^{A}_{\, L} \, g_{KB}
- \delta^{A}_{\, K} \, g_{LB}
\end{equation}
are the analogs of the Lorentz generators $(M_{\mu \nu})^{\alpha}_{\; \beta}
\equiv \delta^{\alpha}_{\, \nu} \, g_{\mu \beta} - \delta^{\alpha}_{\, \mu}
\, g_{\nu \beta}$.

As any other five-vector bivector, $\RR$ can be invariantly decomposed into
a part made only of five-vectors from $\cal Z$ and a part which is the
wedge product of a five-vector from $\cal E$ with some other five-vector.
In the following these two parts of $\RR$ will be called its $\cal Z$- and
$\cal E$-components, respectively. Since $\cal Z$ is isomorphic to $V_{4}$,
at each space-time point to the $\cal Z$-component of $\RR$ one can put
into correspondence a certain four-vector bivector, which, as is seen from
equation (11), is the completely contravariant form of the infinitesimal
rotation four-tensor $\bf R$ introduced above and which for this reason I
will denote with the same letter. It is easy to check that the four-tensor
field $\bf R$ obtained this way is covariantly constant and that if
${\bf E}_{\alpha}$ is some four-vector basis and ${\bf e}_{A}$ is the
associated active regular five-vector basis, then the components of
$\bf R$ in the basis ${\bf E}_{\alpha} \otimes {\bf E}_{\beta}$ equal the
components ${\cal R}^{\alpha \beta}$ of $\RR$ in the basis ${\bf e}_{A}
\otimes {\bf e}_{B}$.

In a similar way, since the maximal vector space of simple bivectors over
$V_{5}$ with the direction vector from $\cal E$ is isomorphic to $V_{4}$,
at each space-time point to the $\cal E$-component of $\RR$ one can put
into correspondence a certain four-vector $\bf A$. For practical reasons,
it is more convenient to establish the isomorphism between the above two
vector spaces not as it has been done in section 3 of part II, but in
a slightly different way: supposing that the space of bivectors is endowed
not with the inner product induced by $h$, but with the inner product
differing from the latter by the factor $\xi^{-1}$. In this case the
components of $\bf A$ in the basis ${\bf E}_{\alpha}$ will equal the
components ${\cal R}^{\alpha 5}$ of $\RR$ in the basis ${\bf e}_{A} \otimes
{\bf e}_{B}$, so $\bf A$ itself will coincide with the infinitesimal
displacement four-vector. At $\bf R \neq 0$ the four-vector field $\bf A$
will not be covariantly constant, which is in agreement with the fact that
the values of $\bf R$ and $\bf A$ at any given point $Q$ determine the
rotation and translation of a Lorentz coordinate system with the origin
at $Q$ that one has to make to perform the considered active Poincare
transformation.

\vspace{3ex} \begin{flushleft}
C. \it Motion of a perfectly rigit body \\
\hspace*{2.5ex} in the language of three-plus-one-vectors
\end{flushleft}
Everything that has been said above about the description of active Poincare
transformations in flat space-time in terms of five-tensors can be applied,
with obvious modifications, to the case of flat three-dimensional Euclidean
space. Instead of five-vectors one should now speak about (3+1)-vectors and
instead of Poincare transformations, about transformations from the group
of motions of three-dimensional Euclidean space. I will now show how the
formalism developed in the previous section can be applied for describing
the motion of a perfectly rigit body in classical nonrelativistic mechanics.
In order not to introduce new notations, I will denote the analogs of
tensors $\TT$ and $\RR$ in three-dimensional space with the same symbols.

Owing to the absolute rigidity of the body in question, its motion can be
viewed as an active transformation of the fields (discrete or continuous)
that describe the distribution of matter inside the body---a transformation
that develops in time. Accordingly, the change in the position of the body
that occurs over a finite time period $t$ can be described invariantly with a
certain (3+1)-tensor, $\TT (t)$, and the rate of this change can be described
with a certain antisymmetric (3+1)-tensor, $\WW$, equal to the ratio of the
(3+1)-tensor $\RR (dt)$ that describes to the first order the infinitesimal
transformation that corresponds to the change in the body position over the
time $dt$ to the magnitude of this time interval:
\begin{equation}
\WW \equiv \RR (dt) / dt.
\end{equation}
In the following, $\WW$ will be referred to as the {\em velocity bivector}
of the body.

As in the case of $\RR$, at every point in space one can put into
correspondence to $\WW$ a pair consisting of a three-vector, which will be
denoted as $-\vec{V}$, and of a three-vector bivector. Since the space is
three-dimensional, it is more convenient to deal not with the latter bivector
itself, but with the three-vector dual to it, which I will denote as
$\vec{\Omega}$. If $\vec{e}_{\Theta}$ is some active regular basis of
(3+1)-vectors and $\vec{E}_{i}$ is the associated three-vector basis, then
the components of $\vec{V}$ and $\vec{\Omega}$ in the latter are related
to the components of $\WW$ in the basis $\vec{e}_{\Theta} \otimes
\vec{e}_{\Sigma}$ in the following way:
\begin{equation} \left\{ \begin{array}{rcccl}
{\cal W}^{\, 5i} & = & - \, {\cal W}^{\, i5} & = & V^{i} \\ {\cal W}^{\, ij}
& = & - \, {\cal W}^{\, ji} & = & \epsilon^{ij}_{\; \; k} \, \Omega^{k} .
\end{array} \right. \end{equation}
It is not difficult to show that at each point in space $\vec{V}$ coincides
with the translational velocity of a frame rigitly fixed to the body and with
the origin at that point. Similarly, $\vec{\Omega}$ can be shown to coincide
with the angular velocity of this frame.

Let us introduce in space an arbitrary system of Cartesian coordinates and
consider the vectors $\vec{V}$ and $\vec{\Omega}$ at the point with
coordinates $x^{i}$. Since the field $\RR$ is covariantly constant, so is
the field $\WW$, and consequently the value of $\WW$ at the considered point
can be obtained by transporting parallelly to this point the value of $\WW$
at the origin. By using formulae (2) one can easily find that in the
$O$-basis associated with the selected coordinates,
\begin{displaymath} \left\{ \begin{array}{rcl}
{\cal W}^{\, 5j}(x) & = & {\cal W}^{\, 5j}(0) + x_{i} \, {\cal W}^{\, ij}
(0) \\ {\cal W}^{\, ij}(x) & = & {\cal W}^{\, ij}(0) ,
\end{array} \right. \end{displaymath}
and substituting the components of $\vec{V}$ and $\vec{\Omega}$ for those of
$\WW$, one obtains
\begin{displaymath} \left\{ \begin{array}{rcl}
V^{i}(x) & = & V^{i}(0) + \epsilon^{i}_{\; jk} \, \Omega^{j}(0) \, x^{k} \\
\Omega^{i}(x) & = & \Omega^{i}(0) .
\end{array} \right. \end{displaymath}
In view of the meaning the vectors $\vec{V}$ and $\vec{\Omega}$ have at a
given point, from the latter formulae follows the well-known rule for
transformation of translational and angular veclocities as one transfers
the origin of the moving frame for which they are defined to another point:
\begin{equation}
\vec{V}' = \vec{V} + \vec{\Omega} \times \vec{X} \; \; \mbox{ and } \; \;
\vec{\Omega}' = \vec{\Omega},
\end{equation}
where $\vec{X}$ is the position vector that connects the old origin with the
new one.

Let us now suppose that there is a particle of the body at the point with
coordinates $x^{i}$. Since at any given moment, the velocity of this particle
coincides with the translational veclocity of the frame connected to the
body with the origin at that point, from equations (17) follows another
well-known relation:
\begin{displaymath}
\vec{v} = \vec{V} + \vec{\Omega} \times \vec{r},
\end{displaymath}
where $\vec{v}$ is the particle velocity and $\vec{r}$ is its position vector
relative to the frame for which $\vec{V}$ and $\vec{\Omega}$ are defined.

Let us now consider the expression for kinetic energy. As one knows, in
the general case  the latter can be presented as a sum of three terms: $(i)$
a term bilinear in $\vec{V}$ and independent of $\vec{\Omega}$, $(ii)$ a term
bilinear in $\vec{\Omega}$ and independent of $\vec{V}$, and $(iii)$ a term
linear both in $\vec{V}$ and in $\vec{\Omega}$. Since with respect to the
three-dimensional space the kinetic energy is a scalar, the above means
that in terms of (3+1)-tensors it can be presented in the following form:
\begin{equation}
E_{\rm kin.} = {\scriptstyle \frac{1}{2}} \, {\cal I}_{\Gamma \Delta \Theta
\Sigma} \, {\cal W}^{\, |\Gamma \Delta |} {\cal W}^{\, |\Theta \Sigma |},
\end{equation}
where ${\cal I}_{\Gamma \Delta \Theta \Sigma}$ are components of some
(3+1)-tensor of rank 4, which by definition have the following symmetry
properties:
\begin{displaymath}
{\cal I}_{\Gamma\Delta\Theta\Sigma} = {\cal I}_{\Theta\Sigma\Gamma\Delta}
\end{displaymath}
and
\begin{displaymath}
{\cal I}_{\Gamma \Delta \Theta \Sigma} = - \, {\cal I}_{\Delta \Gamma
\Theta \Sigma} = - \, {\cal I}_{\Gamma \Delta \Sigma \Theta}.
\end{displaymath}
It is obvious that the kinetic energy of a single particle can be presented
in the same form. By comparing the right-hand side of formula (18) with the
usual expression for the kinetic energy of a point particle in classical
nonrelativistic mechanics, and considering that at the point where the
particle is located, $\vec{V}$ coincides with the particle velocity vector,
one finds that in this case
\begin{equation}
{\cal I}_{\, 5i5j} = - \, {\cal I}_{\, i55j} = - \, {\cal I}_{\, 5ij5} =
{\cal I}_{\, i5j5} = m \cdot \delta_{ij},
\end{equation}
where $m$ is the particle mass, and all other components of $\II$ are zero
(here and below I omit the indices that numerate the particles). To express
the kinetic energy of the body as a whole in terms of the velocities
$\vec{V}$ and $\vec{\Omega}$ that correspond to some moving frame with the
origin at point $O$, let us make use of the fact that the contraction of
(3+1)-tensors is conserved by parallel transport, so the kinetic energy of a
given particle of the body equals the contraction of two samples of tensor
$\WW$ at $O$ with the tensor $\II$ corresponding to this particle,
transported from the point where the particle is located to $O$. Thus, for
every particle
\begin{equation} \left. \begin{array}{l}
E_{\rm kin.} = {\scriptstyle \frac{1}{2}} \,
(\II^{\, \rm transported})_{\, 5i5j} \, V^{i} \, V^{j} \\ \hspace{7ex} + \;
(\II^{\, \rm transported})_{\, 5ijk} \, V^{i} \,
\epsilon^{|jk|}_{\hspace{2.5ex} l} \, \Omega^{l} \\ \hspace{5.5ex} + \;
{\scriptstyle \frac{1}{2}} \, (\II^{\, \rm transported})_{\, ijkl} \,
\epsilon^{|ij|}_{\hspace{1.5ex} m} \, \epsilon^{|kl|}_{\hspace{2ex} n}
\, \Omega^{m} \, \Omega^{n},
\end{array} \right. \end{equation}
and the kinetic energy of the body as a whole is the sum of the expressions
in the right-hand side of this formula, taken over all the particles. If the
considered particle is located at the point with coordinates $x^{i}$, then
by using formulae (2) one can easily find that
\begin{equation} \left. \begin{array}{l}
(\II^{\, \rm transported})_{\, 5i5j} = m \, \delta_{ij} \\ (\II^{\, \rm
transported})_{\, 5ijk} \, \epsilon^{|jk|}_{\hspace{2.5ex} l} = m \,
\epsilon_{il}^{\hspace{1ex} j} x_{j} \\ (\II^{\, \rm transported})_{\, ijkl}
\, \epsilon^{|ij|}_{\hspace{2ex} m} \, \epsilon^{|kl|}_{\hspace{2ex} n} \\
\hspace{18ex} = m \, (\delta_{mn} \, x_{i} x^{i} - x_{m} x_{n}),
\end{array} \right. \end{equation}
and substituting these expressions into formula (20), one obtains the usual
expression for the kinetic energy of a rigit body in terms of $\vec{V}$ and
$\vec{\Omega}$:
\begin{displaymath}
E_{\rm kin.} = {\scriptstyle \frac{1}{2}} \, (\sum m) \, (\vec{V})^{2} +
\vec{V} \cdot \vec{\Omega} \times (\sum m \vec{r}) + {\scriptstyle
\frac{1}{2}} \, I_{ij} \, \Omega^{i} \, \Omega^{j},
\end{displaymath}
where the sum goes over all the particles and
\begin{displaymath}
I_{ij} \equiv \sum m \, (\delta_{ij} \, x_{k} x^{k} - x_{i} x_{j}).
\end{displaymath}
Thus, the moments of inertia of the body with respect to $O$ can be found
by dualizing the $ijkl$-components of the summary (3+1)-tensor $\sum \II$ at
$O$ with respect to the indices $i$ and $j$ and with respect to the indices
$k$ and $l$ by using the {\em three}-dimensional $\epsilon$ tensor. To find
the moments of inertia relative to any other point $O'$, one should simply
transport the tensor $\sum \II$ from $O$ to $O'$ according to the rules of
parallel transport for (3+1)-tensors and dualize its $ijkl$-components at
that point.

Tensor $\II$ can be contracted with only one sample of tensor $\WW$. One
will then obtain a completely covariant antisymmetric (3+1)-tensor of rank
two, which I will denote as $\MM$, with the components
\begin{equation}
{\cal M}_{\Gamma \Delta} \; = \; {\cal I}_{\Gamma \Delta \Theta \Xi} \,
{\cal W}^{\, |\Theta \Xi|}.
\end{equation}
As in the case of $\RR$ and $\WW$, at each point in space one can put into
correspondence to $\MM$ a certain pair consisting of a three-vector 1-form
and a three-vector 2-form. For practical reasons it is convenient first to
replace the 2-form with the 1-form dual to it, thereby obtaining instead of
the original pair a pair consisting of two three-vector 1-forms, and then
replace these latter 1-forms with the corresponding three-vectors.
As a result, to $\MM$ there will correspond a pair consisting of two
three-vectors, which I will denote as $-\vec{P}$ and $\vec{M}$. If
$\widetilde{o}^{\Sigma}$ is the basis of three-plus-one-vector 1-forms dual
to the basis $\vec{e}_{\Theta}$ introduced above, then the components of
$\vec{P}$ and $\vec{M}$ in the associated three-vector basis $\vec{E}_{i}$
will be related to the components of $\MM$ in the basis
$\widetilde{o}^{\Theta} \otimes \widetilde{o}^{\Sigma}$ as follows:
\begin{displaymath}
P^{i} = \delta^{ij} {\cal M}_{5j} \; \; \mbox{ and } \; \; M^{i} =
{\scriptstyle \frac{1}{2}} \, \epsilon^{ijk} {\cal M}_{jk}.
\end{displaymath}

Let us now calculate $\MM$ for a single point particle. According to
equations (19), at the point where the latter is located
\begin{displaymath}
{\cal M}_{5i} = m \, \delta_{ij} v^{j} = m v_{i} \; \; \mbox{ and } \; \;
{\cal M}_{ij} = 0,
\end{displaymath}
so in this case $\vec{P}$ coincides with the particle momentum three-vector,
and $\vec{M}=0$. Let us now suppose that in some system of Cartesian
coordinates the considered particle has the coordinates $x^{i}$. Let us
transport $\MM$ to the origin of this system. Then, according to formulae
(2), in the $O$-basis associated with these coordinates,
\begin{displaymath} \left. \begin{array}{rcl}
{\cal M}_{5j}(0) & = & {\cal M}_{5j}(x) = m v_{j} \\
{\cal M}_{ij}(0) & = & x_{i} {\cal M}_{5j}(x) + x_{j} {\cal M}_{i5}(x) \\
& = & m \, (x_{i} v_{j} - x_{j} v_{i}),
\end{array} \right. \end{displaymath}
and consequently the pair $(-\vec{P},\vec{M})$ corresponding to the
transported $\MM$ is such that $\vec{P}$ coincides with the particle
momentum three-vector transported to the origin according to ordinary rules
of parallel transport for three-vectors, and $\vec{M}$ coincides with the
three-vector of the particle angular momentum relative to the origin. Since
the latter can be selected arbitrarily, this correspondence between $\MM$
and the particle momentum and angular momentum will exist at every point.
Naturally, each of the three-vectors in the pair $(-\vec{P},\vec{M})$ can
be transported to any other point in space according to the rules of
parallel transport for three-vectors, however, the pair as a whole will
correspond to the (3+1)-tensor $\MM$ only at the point with respect to
which the angular momentum is defined.

We thus see that the momentum and angular momentum in classical
nonrelativistic mechanics can be described by a single geometric object---by
the antisymmetric (3+1)-tensor $\MM$. It is natural to called the latter
the {\em momentum--angular momentum tensor}. One of the advantages of such
a description compared to the description in terms of three-vectors is that
$\MM$ can be defined in a purely {\em local} way, with no reference to any
other point in space. For example, in the case of a single point particle,
at the point where the latter is located the $\widetilde{\cal E}$-component
of $\MM$ is expressed in terms of the particle momentum and the
$\widetilde{\cal Z}$-component of $\MM$ is zero. Having defined the
tensor $\MM$ this way, one can then transport it to any other point in
space. This transport will result in that $\MM$ will acquire a nonzero
$\widetilde{\cal Z}$-component, which will be exactly the angular momentum
of the particle relative to the point where $\MM$ has been transported to.

In order to calculate the momentum--angular momentum tensor for a {\em
system} of point particles, one should first transport the tensors $\MM$
corresponding to all the particles to one point in space. It is at this
stage that the tensors $\MM$ of individual particles will acquire nonzero
$\widetilde{\cal Z}$-components, which, when summed up, will give the total
angular momentum of the system relative to the selected point. In the
particular case where the system is a rigit body, its total momentum and
total angular momentum can be expressed in terms of the velocities $\vec{V}$
and $\vec{\Omega}$ corresponding to some frame connected to the body, with
the origin at point $O$. To do this, one should follow the same procedure
that has been used above for calculating the kinetic energy: one should
transport the tensors $\II$ corresponding to individual particles of the
body to $O$, sum them up there, and then contract the sum $\sum \II$ with
the velocity bivector of the body at that point. As a result, one will
obtain the usual expressions for momentum and angular momentum of the rigit
body in terms of $\vec{V}$ and $\vec{\Omega}$, which I will not present here.

Let us now discuss the equations of motion. In the case of a single point
particle one has:
\begin{equation}
d \vec{P} / dt = \vec{F},
\end{equation}
where $\vec{P}$ is the particle momentum and $\vec{F}$ is the acting
force. Since $\vec{P}$ corresponds to the $\widetilde{\cal E}$-component
of the (3+1)-tensor $\MM$, one may suppose that the above equation
of motion corresponds to the $\widetilde{\cal E}$-component of some
three-plus-one-tensor equation. One should expect that the left-hand side
of this latter equation is the time derivative of $\MM$ and that its
right-hand side is some antisymmetric (3+1)-tensor of rank 2, which I will
denote as $\KK$. Thus, the three-plus-one-tensor equation will have the form
\begin{equation}
d \MM / dt = \KK,
\end{equation}
and now we should determine how $\KK$ is related to the known three-vector
quantities.

Let us introduce in space some system of Cartesian coordinates and let
$x(t)$ denote the trajectory of the particle. To evaluate the time derivative
in the left-hand side of equation (24), one should take the tensor
$\MM(t+dt)$ at the point $x(t+dt)$, transport it according to the rules of
parallel transport for (3+1)-tensors to the point $x(t)$, subtract from
it the tensor $\MM (t)$ at that point, and divide the difference by $dt$.
Following this procedure, one will find that in the $O$-basis associated
with the selected coordinates,
\begin{displaymath}
{\cal K}_{5i}(t) = \frac{mv_{i}(t+dt) - mv_{i}(t)}{dt}
= \frac{d(mv_{i})}{dt}  = \delta_{ij} F^{j}(t).
\end{displaymath}
Similarly, since at any $t$ the $\widetilde{\cal Z}$-component of $\MM(t)$
at the point $x(t)$ is zero, one will have
\begin{displaymath} \left. \begin{array}{rcl}
{\cal K}_{ij}(t) & = & v_{i}(t) \, {\cal M}_{5j}(t) + v_{j}(t) \,
{\cal M}_{i5}(t) \\ & = & m v_{i}(t) v_{j}(t) - m v_{j}(t) v_{i}(t) = 0.
\end{array} \right. \end{displaymath}

As $\MM$, $\KK$ can be represented by a pair of three-vectors. The results
we have just obtained mean that at the point where the particle is located,
the pair corresponding to $\KK$ is $(-\vec{F},\vec{0})$. In the particular
case where no forces act on the particle, one obtains
\begin{displaymath}
d \MM / dt = 0,
\end{displaymath}
which is nothing but the conservation law for momentum and angular momentum
of a free particle, written down in the language of (3+1)-tensors.

Suppose now that $\vec{F} \neq 0$. Let us transport the tensors in both sides
of equation (24) to some other point $O$. As we know, the pair corresponding
to the transported $\MM$ will consist of the particle momentum three-vector
with the minus sign and of the three-vector of particle angular momentum
relative to $O$. Similarly, one can find that the pair corresponding to the
transported $\KK$ will consist of the three-vector $-\vec{F}$ transported to
$O$ according to the rules of parallel transport for three-vectors, and of
the three-vector $\vec{K}$ of the force moment relative to $O$. Thus, when
transported to the indicated point, the three-plus-one-tensor equation (24)
is equivalent to the following two three-vector equations: equation (23)
and the equation
\begin{displaymath}
d \vec{M} / dt = \vec{K}.
\end{displaymath}

In the case of a system of point particles, one can sum up equations (24)
corresponding to all the particles in the system, provided one first
transports them all to some point $O$, and obtain the three-plus-one-tensor
equation
\begin{equation}
d \MM^{tot} / dt = \KK^{tot},
\end{equation}
which is apparently equivalent to two three-vector equations that equate
the time derivatives of the total momentum three-vector and of the
three-vector of total angular momentum relative to $O$ respectively to
the three-vector of total force evaluated in the usual way and to the
three-vector of the total force moment relative to $O$.

\vspace{3ex} \begin{flushleft}
D. \it Bivector derivative
\end{flushleft}
Let us consider the group of active Poincare tranformations of scalar,
four-vector and four-tensor fields in flat space-time. Let us distinguish
in it some one-parameter family $\cal H$ that includes the identity
transformation. Let us denote the parameter of this family as $s$ and
the image of an arbitrary field $\cal G$ under a transformation from
$\cal H$ as ${\bf \Pi}_{s} \{ {\cal G} \}$. It is convenient to take
that the identity transformation corresponds to $s=0$.

For the selected one-parameter family $\cal H$ and for any sufficiently
smooth field $\cal G$ from the indicated class of fields, one can define
the derivative
\begin{equation}
{\sf D}_{\cal H}{\cal G} \equiv (d/ds){\bf \Pi}_{s} \{ {\cal G} \}|_{s=0},
\end{equation}
which is a field of the same type as $\cal G$. It is apparent that for
every type of fields, the operators ${\sf D}_{\cal H}$ corresponding to
all possible one-parameter families $\cal H$ make up a 10-dimensional real
vector space, which is nothing but the representation of the Lie algebra of
the Poincare group that corresponds to the considered type of fields.

Let us introduce in space-time some system of Lorentz coordinates
$x^{\alpha}$ and select a basis in the space of operators ${\sf D}_{\cal H}$
consisting of the six operators ${\sf M}_{\mu \nu}$ that correspond to
rotations in the planes $x^{\mu} x^{\nu}$ ($\mu < \nu$) and of the four
operators ${\sf P}_{\! \mu}$ that correspond to translations along the
coordinate axes. If one parametrizes the indicated transformations with the
parameters $\omega^{\alpha \beta}$ and $a^{\alpha}$ introduced in section B,
then for an arbitrary scalar function $f$ one will have
\begin{equation} \begin{array}{lcl}
{\sf P}_{\! \mu}f(x) & = & \partial_{\mu}f(x) \\ {\sf M}_{\mu \nu} f(x)
& = & x_{\nu} \partial_{\mu} f(x) - x_{\mu} \partial_{\nu} f(x),
\end{array} \end{equation}
for an arbitrary four-vector field $\bf U$ one will have
\begin{equation} \begin{array}{lcl}
({\sf P}_{\! \mu}{\bf U})^{\alpha}(x) & = & \partial_{\mu}
U^{\alpha}(x) \\ ({\sf M}_{\mu \nu}{\bf U})^{\alpha}(x) & = &
x_{\nu} \partial_{\mu} U^{\alpha}(x)  - x_{\mu} \partial_{\nu} U^{\alpha}(x)
\\ & & \hspace{7ex} + \; (M_{\mu \nu})^{\alpha}_{\; \beta} \, U^{\beta}(x),
\end{array} \end{equation}
where the components correspond to the Lorentz four-vector basis associated
with the selected coordinates; and so on.

With transition to some other system of Lorentz coordinates with the origin
at the {\em same} point, the derivatives ${\sf M}_{\mu \nu}{\cal G}$
transform with respect to the indices $\mu$ and $\nu$ as components of a
four-vector 2-form, and the derivatives ${\sf P}_{\! \mu}{\cal G}$ transform
with respect to $\mu$ as components of a four-vector 1-form. Consequently,
if one constructs out of these quantities the fields
\begin{equation}
{\sf P}{\cal G} \equiv {\sf P}_{\! \mu}{\cal G} \cdot \widetilde{\bf O}^{\mu}
\; \mbox{ and } \; {\sf M}{\cal G} \equiv {\sf M}_{|\mu \nu|}{\cal G} \cdot
\widetilde{\bf O}^{\mu} \wedge \widetilde{\bf O}^{\nu},
\end{equation}
where $\widetilde{\bf O}^{\mu}$ is the basis of four-vector 1-forms
associated with the selected coordinate system, these fields will be the
same at any choice of the latter. From definition (29) it follows that at
every point in space-time
\begin{equation} \begin{array}{c}
{\sf P}_{\! \mu}{\cal G} = \; < {\sf P}{\cal G} \, , \, {\bf E}_{\mu}> \\
{\sf M}_{\mu \nu}{\cal G} = \; < {\sf M} {\cal G} \, , \, {\bf E}_{\mu}
\wedge {\bf E}_{\nu} >,
\end{array} \end{equation}
where ${\bf E}_{\mu}$ is the four-vector basis corresponding to the Lorentz
coordinate system with respect to which the operators ${\sf P}_{\! \mu}
{\cal G}$ and ${\sf M}_{\mu \nu}{\cal G}$ are defined. If the fields
${\sf P}{\cal G}$ and ${\sf M}{\cal G}$ were completely independent of the
choice of the Lorentz coordinate system, then basing on relations (30) one
could regard ${\sf M}_{\mu \nu}{\cal G}$ as a special kind of derivative
whose argument is a four-vector bivector:
\begin{displaymath}
{\sf M}_{\mu \nu}{\cal G} \equiv {\sf D}_{{\bf E}_{\mu} \wedge {\bf E}_{\nu}}
{\cal G},
\end{displaymath}
and ${\sf P}_{\! \mu}{\cal G}$ as a derivative whose argument is a
four-vector:
\begin{displaymath}
{\sf P}_{\! \mu}{\cal G} \equiv {\sf D}_{{\bf E}_{\mu}}{\cal G},
\end{displaymath}
and then at any point in space-time, the family $\cal H$ for which the
derivative ${\sf D}_{\cal H}$ is evaluated could be identified in a
coordinate-free way by indicating the four-vector and the four-vector
bivector that correspond to this family. The field ${\sf P}{\cal G}$ is
indeed independent of the choice of the coordinate system, and it is easy
to see that for all types of fields from the considered class of fields
the operator ${\sf D}_{{\bf E}_{\mu}}$ coincides with the operator of the
covariant derivative in the direction of the four-vector ${\bf E}_{\mu}$
(I am talking about flat space-time only). However, the field
${\sf M}{\cal G}$ does depend on the choice of the origin, since under the
translation $x^{\alpha} \rightarrow x^{\alpha} + a^{\alpha}$ it transforms
as
\begin{displaymath}
{\sf M}{\cal G} \rightarrow {\sf M}{\cal G} + {\sf P}{\cal G} \wedge
\widetilde{\bf A},
\end{displaymath}
where $\widetilde{\bf A} \equiv a_{\alpha} \widetilde{\bf O}^{\alpha}$,
so one cannot regard ${\sf M}_{\mu \nu}{\cal G}$ as a derivative whose
argument is a four-vector bivector irrespective of the choice of the
coordinate system.

Since derivative (26) is associated with active Poincare transformations,
basing on the results of section B one may expect that at any point in
space-time the derivatives ${\sf D}_{\cal H}$ corresponding to various
one-parameter families $\cal H$ can be parametrized invariantly with
five-vector bivectors. To see that this is indeed so, one may observe
that with transition from one Lorentz coordinate system to another, the
quantities ${\sf M}_{\mu \nu}{\cal G}$ and ${\sf P}_{\! \mu}{\cal G}$
transform respectively as the $\mu \nu$- and $\mu 5$-components of a
five-vector 2-form in the $P$-basis. Consequently, the field
\begin{displaymath}
{\sf D}{\cal G} \equiv {\sf M}_{|\mu \nu|}{\cal G} \cdot
\widetilde{\bf q}^{\mu} \wedge \widetilde{\bf q}^{\nu} + {\sf P}_{\! \mu}
{\cal G} \cdot \widetilde{\bf q}^{\mu} \wedge \widetilde{\bf q}^{5},
\end{displaymath}
where $\widetilde{\bf q}^{A}$ is the basis of five-vector 1-forms dual to
the $P$-basis, ${\bf p}_{A}$, associated with the selected Lorentz coordinate
system, will be the same at any choice of the latter. Similar to equations
(30), one will have the relations
\begin{displaymath} \begin{array}{c}
{\sf P}_{\! \mu}{\cal G} = \; < {\sf D}{\cal G} \, , \, {\bf p}_{\mu}
\wedge {\bf p}_{5} > \\ {\sf M}_{\mu \nu}{\cal G} = \;
< {\sf D} {\cal G} \, , \, {\bf p}_{\mu} \wedge {\bf p}_{\nu} >,
\end{array} \end{displaymath}
basing on which one can regard ${\sf P}_{\! \mu}{\cal G}$ and
${\sf M}_{\mu \nu}{\cal G}$ as particular values of the derivative whose
argument is a five-vector bivector, and which in view of this I will call
the {\em bivector} derivative. For any Lorentz coordinate system one will
apparently have
\begin{equation}
{\sf P}_{\! \mu}{\cal G} = {\sf D}_{{\bf p}_{\mu} \wedge
{\bf p}_{5}}{\cal G} \; \mbox{ and } \; {\sf M}_{\mu \nu}{\cal G}
= {\sf D}_{{\bf p}_{\mu} \wedge {\bf p}_{\nu}}{\cal G},
\end{equation}
where ${\bf p}_{A}$ is the $P$-basis associated with these coordinates.
Comparing the latter formulae with formulae (27) at the origin, one can see
that for any active regular basis ${\bf e}_{A}$ and any scalar function $f$,
\begin{equation}
{\sf D}_{{\bf e}_{\mu} \wedge {\bf e}_{5}} f = \partial_{{\bf e}_{\mu}} f
\; \mbox{ and } \; {\sf D}_{{\bf e}_{\mu} \wedge {\bf e}_{\nu}} f = 0.
\end{equation}
 From these equations it follows that at the point with coordinates
$x^{\alpha}$,
\begin{displaymath}
{\sf D}_{{\bf p}_{\mu} \wedge {\bf p}_{5}}f = {\sf D}_{{\bf p}^{\cal Z}_{\mu}
\wedge {\bf p}_{5}} f = {\sf D}_{{\bf e}_{\mu} \wedge {\bf e}_{5}} f =
\partial_{{\bf e}_{\mu}} f = \partial_{\mu} f
\end{displaymath}
and
\begin{displaymath} \begin{array}{lcl}
{\sf D}_{{\bf p}_{\mu} \wedge {\bf p}_{\nu}} f & \! = & \!
{\sf D}_{{\bf e}_{\mu} \wedge {\bf e}_{\nu}} f + x_{\nu}
{\sf D}_{{\bf e}_{\mu} \wedge {\bf e}_{5}} f + x_{\mu} {\sf D}_{{\bf e}_{5}
\wedge {\bf e}_{\nu}} f \\ & \! = & \! x_{\nu} \partial_{\mu} f - x_{\mu}
\partial_{\nu} f, \end{array} \end{displaymath}
which is in agreement with formulae (27) in the general case (in the
latter two chains of equations and in equations (33) and (34) that
follow, ${\bf e}_{A}$ denotes the $O$-basis associated with the considered
coordinates). Comparing formulae (31) with formulae (28) at the origin,
one can see that for any Lorentz four-vector basis ${\bf E}_{\alpha}$,
\begin{equation}
{\sf D}_{{\bf e}_{\mu} \wedge {\bf e}_{5}} {\bf E}_{\alpha} = {\bf 0}
\; \mbox{ and } \; {\sf D}_{{\bf e}_{\mu} \wedge {\bf e}_{\nu}}
{\bf E}_{\alpha} = {\bf E}_{\beta} \, (M_{\mu \nu})^{\beta}_{\; \alpha},
\end{equation}
so for any such basis
\begin{equation}
{\sf D}_{{\bf p}_{A} \wedge {\bf p}_{B}} {\bf E}_{\alpha} =
{\sf D}_{{\bf e}_{A} \wedge {\bf e}_{B}} {\bf E}_{\alpha}
\end{equation}
at all $A$ and $B$. From the properties of Poincare transformations
and from definition (26) it follows that for any scalar function $f$
and any four-vector field $\bf V$,
\begin{equation}
{\sf D}_{{\bf p}_{A} \wedge {\bf p}_{B}} (f {\bf V}) = {\sf D}_{{\bf p}_{A}
\wedge {\bf p}_{B}} f \cdot {\bf V} + f \cdot {\sf D}_{{\bf p}_{A} \wedge
{\bf p}_{B}} {\bf V},
\end{equation}
which together with equations (33) and (34) gives formulae (28) for an
arbitrary four-vector field $\bf U$. Similar formulae can be obtained for
all other four-tensor fields.

One can now consider a more general derivative than ${\sf D}_{\cal H}$ by
allowing the one-parameter family $\cal H$ to vary from point to point. For
any field $\cal G$ from the considered class of fields, such a derivative is
a field whose value at each space-time point coincides with the value of one
of the fields ${\sf D}_{\cal H}{\cal G}$ for some family $\cal H$, which at
different points may be different. Everywhere below, when speaking of the
bivector derivative I will refer to this more general type of
differentiation.

According to the results obtained above, any such derivative can be uniquely
fixed by specifying a certain field of five-vector bivectors. Therefore, by
analogy with the covariant derivative, for any type of fields $\sf D$ can
be formally regarded as a map that puts into correspondence to every pair
consisting of a bivector field and a field of the considered type another
field of that type. For example, the bivector derivative for four-vector
fields can be viewed as a map
\begin{equation}
\FF \! \wedge \! \FF \times \DD \rightarrow \DD,
\end{equation}
where $\FF \! \wedge \! \FF$ is the set of all fields of five-vector
bivectors and $\DD$ is the set of all four-vector fields. From the
definition of the bivector derivative it follows that map (36) has the
following formal properties: for any scalar functions $f$ and $g$, any
four-vector fields $\bf U$ and $\bf V$, and any bivector fields $\AAA$
and $\BB$,
\begin{flushright}
\hfill ${\sf D}_{(f{\cal A} + g{\cal B})} {\bf U} = f \cdot {\sf D}_{\cal A}
{\bf U} + g \cdot {\sf D}_{\cal B} {\bf U}$ \hfill {\rm (37a)} \\
\hfill ${\sf D}_{\cal A} ({\bf U + V}) = {\sf D}_{\cal A} {\bf U} +
{\sf D}_{\cal A} {\bf V}$ \hspace*{3.5ex} \hfill {\rm (37b)} \\
\hfill ${\sf D}_{\cal A} (f{\bf U}) = {\sf D}_{\cal A} f \cdot {\bf U} +
f \cdot  {\sf D}_{\cal A} {\bf U}.$ \hspace*{0.5ex} \hfill {\rm (37c)}
\end{flushright} \setcounter{equation}{37}
In the third equation, the action of $\sf D$ on the function $f$ is
determined by the rules:
\begin{equation} \left\{ \begin{array}{l}
{\sf D}_{\cal (A + B)} f = {\sf D}_{\cal A} f + {\sf D}_{\cal B} f, \\
{\sf D}_{\cal A^{Z}} f = 0, \; \; {\sf D}_{\cal A^{E}}f = \partial_{\bf A}f,
\end{array} \right. \end{equation}
where $\bf A$ denotes the four-vector field that corresponds to the
$\cal E$-component of $\AAA$.

The properties of $\sf D$ presented above are similar to the three main
properties of the covariant derivative that are used for defining the
latter formally. Using properties (37) for the same purpose is not very
convenient, since to define the bivector derivative completely one has to
supplement them with the formulae that determine the relation of $\sf D$ to
space-time metric, and usually from such relations one is already able to
derive part of the properties expressed by equations (37). As an example,
let us consider the formulae that express the operator $\sf D$ in terms
of the operator $\dotnabla$ of the torsion-free $g$-conserving covariant
derivative and of the linear local operator $\widehat{\bf M}$ defined below,
both of which are completely determined by the metric. For an arbitrary
four-vector field $\bf U$ one has:
\begin{equation}
{\sf D}_{\cal A^{E}} {\bf U} = \dotnabla_{\bf A} {\bf U} \; \mbox{ and } \;
{\sf D}_{\cal A^{Z}} {\bf U} = \widehat{\bf M}_{\bf B} {\bf U},
\end{equation}
where $\bf A$, as in definition (38), denotes the four-vector field
corresponding to the $\cal E$-compo\-nent of $\AAA$, $\bf B$ denotes the
field of four-vector bivectors corresponding to the $\cal Z$-component of
$\AAA$, and the operator $\widehat{\bf M}$, which depends linearly on its
argument, has the following components in an arbitrary four-vector basis
${\bf E}_{\alpha}$:
\begin{displaymath}
\widehat{\bf M}_{{\bf E}_{\alpha} \wedge {\bf E}_{\beta}} {\bf E}_{\mu}
= {\bf E}_{\nu} (M_{\alpha \beta})^{\nu}_{\, \mu}.
\end{displaymath}
It is easy to see that properties (37b) and (37c) follow from formulae
(39) and property (37a), and property (37a) itself follows from
equations (39) and the following simpler property:
\begin{displaymath}
{\sf D}_{({\cal A + B})} {\bf U} =
{\sf D}_{\cal A} {\bf U} + {\sf D}_{\cal B} {\bf U},
\end{displaymath}
which is similar to the first equation in definition (38) and which,
together with equations (39), can serve as a definition of the bivector
derivative for four-vector fields.

The action of operator $\sf D$ on all other four-tensor fields can be
defined either independent\-ly---according to formula (26), or as in the
case of the covariant derivative---according to the equations that express
the Leibniz rule in application to the contraction of a four-vector 1-form
with a four-vector field and to the tensor product of any two four-tensor
fields. The corresponding formulae are quite obvious and will not be
presented here.

For the bivector derivative one can define the analogs of connection
coefficients. Namely, for any set of basis four-vector fields
${\bf E}_{\alpha}$ and any set of basis five-vector fields ${\bf e}_{A}$
one can take
\begin{equation}
{\sf D}_{AB} {\bf E}_{\mu} = {\bf E}_{\nu} \Gamma^{\nu}_{\; \mu AB},
\end{equation}
where ${\sf D}_{AB} \equiv {\sf D}_{{\bf e}_{A} \wedge {\bf e}_{B}}$.
According to equations (33), for any Lorentz four-vector basis and any
standard five-vector basis associated with it, one has
\begin{equation} \begin{array}{l}
\Gamma^{\mu}_{\; \nu \alpha 5} = - \Gamma^{\mu}_{\; \nu 5 \alpha} = 0 \\
\Gamma^{\mu}_{\; \nu \alpha \beta} = - \Gamma^{\mu}_{\; \nu \beta \alpha}
= (M_{\alpha \beta})^{\mu}_{\; \nu}.
\end{array} \end{equation}
The bivector connection coefficients for any other choice of the basis
fields can be found either by using the following transformation formula:
\begin{equation} \begin{array}{l}
\Gamma'^{\mu}_{\; \; \nu AB} = (\Lambda^{-1})^{\mu}_{\, \sigma}
\Gamma^{\sigma}_{\; \tau ST} \Lambda^{\tau}_{\, \nu} L^{S}_{\, A}
L^{T}_{\, B} \\ \hspace{14ex} + \; (\Lambda^{-1})^{\mu}_{\, \sigma}
({\sf D}_{ST} \Lambda^{\sigma}_{\, \nu}) L^{S}_{\, A} L^{T}_{\, B},
\end{array} \end{equation}
which corresponds to the transformations ${\bf E}'_{\alpha} = {\bf E}_{\beta}
\Lambda^{\beta}_{\, \alpha}$ and ${\bf e}'_{A} = {\bf e}_{B} L^{B}_{\, A}$
of the four- and five-vector basis fields, or by using formulae (39). In
particular, for an arbitrary four-vector basis and the corresponding active
regular five-vector basis one has
\begin{equation}
\Gamma^{\mu}_{\; \nu \alpha 5} = \Gamma^{\mu}_{\; \nu \alpha} \; \mbox{ and }
\; \Gamma^{\mu}_{\; \nu \alpha \beta} = (M_{\alpha \beta})^{\mu}_{\; \nu},
\end{equation}
where $\Gamma^{\mu}_{\; \nu \alpha}$ are ordinary four-vector connection
coefficients associated with $\dotnabla$.

Let me also observe that the quantities ${\sf D}_{AB}{\cal G}$ for an
arbitrary field ${\cal G}$ can be presented as derivatives with respect
to the components of the bivector $\RR$ introduced in section B. Indeed,
according to the definition of the operators ${\sf P}_{\! \mu}$ and
${\sf M}_{\mu \nu}$, one has
\begin{equation} \begin{array}{l}
{\sf D}_{5 \mu}{\cal G} = - (\partial / \partial a^{\mu}) {\bf \Pi} \{
{\cal G} \}|_{\omega^{\alpha \beta} = a^{\alpha} = 0} \\
{\sf D}_{\mu \nu} {\cal G} = (\partial / \partial \omega^{\mu \nu}) {\bf \Pi}
\{ {\cal G} \}|_{\omega^{\alpha \beta}=a^{\alpha}=0} \; \; \;
{\scriptstyle (\mu < \nu)},
\end{array} \end{equation}
where it is assumed that the image ${\bf \Pi} \{ {\cal G} \}$ of field
$\cal G$ under an infinitesimal Poincare transformation is a function of
parameters $\omega^{\alpha \beta}$ and $a^{\alpha}$. In view of equations
(11), the latter formulae can be rewritten as
\begin{equation} \begin{array}{l}
{\sf D}_{5 \mu}{\cal G} = (\partial / \partial {\cal R}^{5 \mu}) {\bf \Pi}
\{ {\cal G} \}|_{{\cal R}^{\alpha \beta} = {\cal R}^{5 \alpha} = 0} \\
{\sf D}_{\mu \nu}{\cal G} = (\partial / \partial {\cal R}^{\mu \nu})
{\bf \Pi} \{ {\cal G} \}|_{{\cal R}^{\alpha \beta}={\cal R}^{5 \alpha}=0},
\end{array} \end{equation}
and since ${\sf D}_{\mu \nu} = - {\sf D}_{\nu \mu}$ and ${\cal R}^{\mu \nu}=
- {\cal R}^{\nu \mu}$, the second equation will be valid at $\mu > \nu$ as
well. Since ${\sf D}_{\mu 5} = - {\sf D}_{5 \mu}$, one can write that
\begin{displaymath}
{\sf D}_{\mu 5}{\cal G} = (\partial / \partial (-{\cal R}^{5 \mu}))
{\bf \Pi} \{ {\cal G} \}|_{{\cal R}^{\alpha \beta}={\cal R}^{5 \alpha}=0},
\end{displaymath}
so if one takes ${\cal R}^{\mu 5} = - {\cal R}^{5 \mu}$, as it has been done
in section B, one will have
\begin{equation}
{\sf D}_{AB}{\cal G} = (\partial / \partial {\cal R}^{AB}) {\bf \Pi}
\{ {\cal G} \}|_{{\cal R}=0},
\end{equation}
for all $A \neq B$, which is one more argument in favour of the choice (12).

\vspace{3ex} \begin{flushleft}
E. \it Bivector derivative of the Lagrange function
\end{flushleft}
In the previous section I have defined the bivector derivative for scalar,
four-vector and four-tensor fields in flat space-time. In a similar manner
the bivector derivative can be defined for more complicated objects. As an
example, I will now consider the definition of the analog of the bivector
derivative in three-dimensional Euclidean space for the Lagrange function of
a system of several point particles, in classical nonrelativistic mechanics.

As is known, the state of motion of such a system at every moment of time
can be fixed by specifying the position of each particle in space and its
velocity. The Lagrange function $\bf L$ for this system can be viewed as a
map that puts into correspondence to each allowed state of motion $\cal C$
a real number, $\bf L (\cal C)$. In the general case this map may be
explicitly time-dependent.

The bivector derivative of the Lagrange function can be defined according to
an equation similar to formula (26). For that one apparently has to define
first how $\bf L$ changes under active transformations from the group of
motions of three-dimensional Euclidean space. This can be done in a standard
way if one knows how such transformations affect the states of motion of the
system. Namely, for any transformation from the indicated group the image
$\bf \Pi \{ L \}$ of the Lagrange function is such a function of the state
of motion that for any $\cal C$
\begin{equation}
\bf \Pi \{ L \} (\Pi \{ \cal C \}) = \bf L(\cal C),
\end{equation}
where $\bf \Pi \{ \cal C \}$ is the image of state $\cal C$ under the
considered active transformation. It is apparent that equation (47) can be
presented in the following equivalent way:
\begin{equation}
\bf \Pi \{ L \} (\cal C) = \bf L(\Pi^{-1} \{ \cal C \}),
\end{equation}
where $\bf \Pi^{-1}$ is the transformation inverse to $\bf \Pi$. Basing
on definition (47), one can define the derivative of $\bf L$ relative
to some one-parameter family of transformations $\cal H$ as follows:
${\sf D}_{\cal H}{\bf L}$ is such a real-valued function of the state of
motion that for any $\cal C$
\begin{equation}
{\sf D}_{\cal H}{\bf L}({\cal C}) = (d/ds){\bf \Pi}_{s} \{ {\bf L} \}
({\cal C})|_{s=0},
\end{equation}
where $s$ is the parameter of the considered family. If one proceeds from
definition (48), then instead of (49) one will have the following equivalent
definition:
\begin{equation}
{\sf D}_{\cal H}{\bf L}({\cal C}) = (d/ds){\bf L}({\bf \Pi}^{-1}_{s}
\{ {\cal C} \} )|_{s=0}.
\end{equation}

By using some particular set of variables for characterizing the state of
motion of the system, for example, the coordinates of all the particles in
some Cartesian coordinate system and the components of their velocities
in the corresponding three-vector basis, it is not difficult to show that
derivatives (49) for all one-parameter families $\cal H$ are correlated
with the derivatives of scalar, three-vector and three-tensor fields
relative to all these families in the following sense: if families $\cal H$,
$\cal H'$, and $\cal H''$ are such that for any field $\cal G$ from the
indicated class of fields,
\begin{displaymath}
{\sf D}_{\cal H}{\cal G} =
a \cdot {\sf D}_{\cal H'}{\cal G} + b \cdot {\sf D}_{\cal H''}{\cal G},
\end{displaymath}
where $a$ and $b$ are some real numbers, then for any state of motion
$\cal C$,
\begin{displaymath}
{\sf D}_{\cal H}{\bf L}({\cal C}) = a \cdot {\sf D}_{\cal H'}
{\bf L}({\cal C}) + b \cdot {\sf D}_{\cal H''}{\bf L}({\cal C}).
\end{displaymath}
This fact enables one to construct another function of the state of motion,
which, as $\bf L$, may explicitly depend on time, but whose values will be
covariantly constant fields of antisymmetric (3+1)-tensors of rank two rather
than numbers. To this end, let us observe that as in the case of space-time,
to every one-parameter family $\cal H$ one can put into correspondence a
certain covariantly constant field $\AAA$ of three-plus-one-vector bivectors,
such that the derivative ${\sf D}_{\cal H}{\cal G}$ of any scalar,
three-vector or three-tensor field $\cal G$ can be presented as the
contraction $< \! {\sf D} {\cal G}, \AAA \! >$, where ${\sf D}{\cal G}$ is
a three-plus-one-vector 2-form independent of $\cal H$, whose values are
quantities of the same kind as those of $\cal G$. In a similar way one can
define the scalar-valued three-plus-one-vector 2-form ${\sf D}{\bf L}$,
which will be a function of the state of motion of the system (and of time),
by taking that for any $\cal C$
\begin{displaymath}
< {\sf D}{\bf L}({\cal C}), \AAA > \; = {\sf D}_{\cal H}{\bf L}({\cal C}),
\end{displaymath}
where $\AAA$ is an arbitrary covariantly constant field of
three-plus-one-vector bivectors and $\cal H$ is the one-parameter family
that corresponds to it. Since ${\sf D}_{\cal H}{\bf L}({\cal C})$ is simply
a number, from the fact that $\AAA$ is covariantly constant follows that
at any $\cal C$ the field ${\sf D}{\bf L}({\cal C})$ will be covariantly
constant, too.

Characterizing the state of motion of the system with coordinates
$x^{i}_{\ell}$ of all the particles in some Cartesian coordinate system
and with the components $v^{i}_{\ell} \equiv dx^{i}_{\ell}/dt$ of their
velocities in the corresponding three-vector basis (the index $\ell$
numerates the particles), it is not difficult to evaluate the components
of ${\sf D} {\bf L}$ in the basis of three-plus-one-vector 2-forms
corresponding to the $P$-basis of (3+1)-vectors associated with the
selected coordinates. One obtains:
\begin{equation}
{\sf D}_{i5}{\bf L} = \sum_{\ell} \partial {\bf L}/\partial x^{i}_{\ell}
\end{equation}
and
\begin{equation}
{\sf D}_{ij}{\bf L} = \sum_{\ell} \left(
x_{j \, \ell} \cdot \partial {\bf L}/\partial x^{i}_{\ell} -
x_{i \, \ell} \cdot \partial {\bf L}/\partial x^{j}_{\ell} \right),
\end{equation}
where $x_{i \, \ell} \equiv \delta_{ij} x^{j}_{\ell}$. Since the quantities
$\partial {\bf L}/\partial x^{i}_{\ell}$ are covariant components of the
force that acts on the $\ell${\em th} particle relative to the basis of
three-vector 1-forms associated with the selected coordinate system,
equations (51) and (52) mean that at the origin $O$ of this system the
$\widetilde{\cal E}$-component of ${\sf D}{\bf L}$ corresponds to the
three-vector 1-form of the total force that acts on the system and the
$\widetilde{\cal Z}$-component of ${\sf D}{\bf L}$ corresponds to the
three-vector 2-form of the total force moment relative to $O$ taken
with the opposite sign. Therefore, $-{\sf D}{\bf L}$ is exactly the
three-plus-one-vector 2-form $\KK^{tot}$ introduced in section C.

Among other thing, from the latter fact follows the result we have obtained
earlier: that momentum and angular momentum of a system of particles in
classical nonrelativistic mechanics can be described by a single local
object---by a three-plus-one-vector 2-form. Indeed, according to the
equations of motion, the force that acts on the particle and the moment of
this force relative to an arbitrary point $O$ are total time derivatives
respectively of the particle momentum and of its angular moment relative to
$O$. Consequently, the three-plus-one-vector 2-form $\KK$ is also a total
time derivative of some (3+1)-tensor. Since in the nonrelativistic case time
is an external parameter, the rank of this (3+1)-tensor should be the same
as that of $\KK$, and according to what has been said above, at an arbitrary
point in space the $\widetilde{\cal E}$-component of this (3+1)-tensor
will correspond, with the opposite sign, to the three-vector 1-form of the
particle momentum, and its $\widetilde{\cal Z}$-component will correspond to
the three-vector 2-form of the particle angular momentum  relative to that
point.

\vspace{3ex} \begin{flushleft}
\bf Acknowledgements
\end{flushleft}
I would like to thank V. D. Laptev for supporting this work. I am grateful
to V. A. Kuzmin for his interest and to V. A. Rubakov for a very helpful
discussion and advice. I am indebted to A. M. Semikhatov of the Lebedev
Physical Institute for a very stimulating and pleasant discussion and to
S. F. Prokushkin of the same institute for consulting me on the Yang-Mills
theories of the de Sitter group. I would also like to thank L. A. Alania,
S. V. Aleshin, and A. A. Irmatov of the Mechanics and Mathematics Department
of the Moscow State University for their help and advice.

\vspace{3ex} \begin{flushleft} \bf
Appendix: Contravariant basis
\end{flushleft}
As in the case of any other vector space endowed with a nondegenerate inner
product, to any basis ${\bf E}_{\alpha}$ in $V_{4}$ one can put into
correspondence the basis
\begin{equation}
{\bf E}^{\alpha} \equiv g^{\alpha \beta} {\bf E}_{\beta},
\end{equation}
which will be called {\em contravariant}. Here, as usual, $g^{\alpha \beta}$
denote the matrix inverse to $g_{\alpha \beta} \equiv g({\bf E}_{\alpha},
{\bf E}_{\beta})$. Definition (53) is equivalent to the following relation:
\begin{displaymath}
< \widetilde{\bf O}^{\alpha},{\bf V} > \; = g({\bf E}^{\alpha},{\bf V})
\; \mbox{ for any four-vector } {\bf V},
\end{displaymath}
where $\widetilde{\bf O}^{\alpha}$ is the basis of four-vector 1-forms
dual to ${\bf E}_{\alpha}$. The latter relation means that the
four-vectors ${\bf E}^{\alpha}$ are inverse images of the basis 1-forms
$\widetilde{\bf O}^{\alpha}$ under the map $V_{4} \rightarrow
\widetilde{V}_{4}$ defined by the inner product $g$. From definition
(53) it also follows that $g({\bf E}_{\alpha}, {\bf E}^{\beta}) =
\delta_{\alpha}^{\, \beta}$, so the four-vector 1-forms
$\widetilde{\bf O}_{\alpha}$ that make up the basis dual to
${\bf E}^{\alpha}$ are images of the basis four-vectors ${\bf E}_{\alpha}$
under the indicated map. It is evident that when ${\bf E}_{\alpha}$ is a
Lorentz basis in flat space-time, associated with some Lorentz coordinate
system $x^{\alpha}$, one has ${\bf E}^{\alpha} = \partial / \partial
x_{\alpha}$, where $x_{\alpha}$ are the corresponding covariant coordinates.
It is also evident that at any affine connection relative to which the
metric tensor is covariantly constant, the relation between the bases
${\bf E}_{\alpha}$ and ${\bf E}^{\alpha}$ is preserved by parallel
transport, and it is easy to see that in this case
\begin{displaymath}
\nabla_{\mu} {\bf E}^{\alpha} = - \; \Gamma^{\alpha}_{\; \beta \mu}
{\bf E}^{\beta},
\end{displaymath}
where $\Gamma^{\alpha}_{\; \beta \mu}$ are the connection coefficients
corresponding to the basis fields ${\bf E}_{\alpha}$.

In a similar manner, by using the nondegenerate inner product $h$, one can
define the contravariant basis ${\bf e}^{A}$ corresponding to an arbitrary
five-vector basis ${\bf e}_{A}$:
\begin{displaymath}
< \widetilde{\bf o}^{A},{\bf v} > \; = h({\bf e}^{A},{\bf v})
\; \mbox{ for any five-vector } {\bf v},
\end{displaymath}
where $\widetilde{\bf o}^{A}$ is the basis of five-vector 1-forms dual
to ${\bf e}_{A}$. However, such a definition of the contravariant basis
is inconvenient in two ways. First of all, the contravariant basis
corresponding to an arbitrary standard basis ${\bf e}_{A}$ will in general
not be a standard basis itself (this will be the case only if ${\bf e}_{A}$
is a {\em regular} basis). Secondly, since the inner product $h$ is
not conserved by parallel transport, the latter will not preserve the
correspondence between ${\bf e}_{A}$ and ${\bf e}^{A}$ either. In view
of this, in the case of five-vectors it is more convenient to define
the contravariant basis in another way. Namely, for any standard basis
${\bf e}_{A}$ one takes that
\begin{equation}
{\bf e}^{\alpha} = g^{\alpha \beta} {\bf e}_{\beta} \; \mbox{ and } \;
{\bf e}^{5} = {\bf e}_{5}.
\end{equation}
It is a simple matter to see that the first four vectors of the contravariant
basis defined this way satisfy the relation
\begin{displaymath}
< \widetilde{\bf o}^{\alpha},{\bf v} > \; = g({\bf e}^{\alpha},{\bf v})
\; \mbox{ for any five-vector } {\bf v},
\end{displaymath}
which, however, is not equivalent to the first equation in definition
(54) since it does not fix the $\cal E$-components of the vectors
${\bf e}^{\alpha}$. It is not difficult to see that the basis of
five-vector 1-forms dual to ${\bf e}^{A}$, which will be denoted as
$\widetilde{\bf o}_{A}$, is expressed in terms of the basis 1-forms
$\widetilde{\bf o}^{A}$ as
\begin{displaymath}
\widetilde{\bf o}_{\alpha} = \widetilde{\bf o}^{\beta} g_{\beta \alpha}
\; \mbox{ and } \; \widetilde{\bf o}_{5} = \widetilde{\bf o}^{5}.
\end{displaymath}
The first of these equations is equivalent to the relation
\begin{displaymath}
< \widetilde{\bf o}_{\alpha},{\bf v} > \; = g({\bf e}_{\alpha},{\bf v})
\; \mbox{ for any five-vector } {\bf v},
\end{displaymath}
which means that $\widetilde{\bf o}_{\alpha}$ are images of the basis
five-vectors ${\bf e}_{\alpha}$ with respect to the map $\vartheta_{g}:
V_{5} \rightarrow \widetilde{V}_{5}$ defined in section 3 of part II.
It is also evident that from ${\bf e}_{\alpha} \in {\bf E}_{\alpha}$
follows ${\bf e}^{\alpha} \in {\bf E}^{\alpha}$.

According to definition (54), the contravariant $P$-basis associated
with some system of Lorentz coordinates $x^{\alpha}$ in flat space-time
is expressed in terms of the corresponding contravariant $O$-basis as
\begin{equation}
{\bf p}^{\alpha} = {\bf e}^{\beta} + x^{\alpha} {\bf e}^{5}
\; \mbox{ and } \; {\bf p}^{5} = {\bf e}^{5}.
\end{equation}
Among other things, from the latter equation it follows that in such a
$P$-basis, the five-vector 1-form $\widetilde{\bf x}$ introduced in section
5 of part I has the components $(x^{\alpha},1)$. Since the correspondence
between ${\bf p}_{A}$ and ${\bf p}^{A}$ is preserved by parallel
transport, from formula (55) one finds that for the contravariant $O$-basis
\begin{displaymath}
\nabla_{\mu} {\bf e}^{\alpha} = - \, \delta_{\mu}^{\alpha} {\bf e}^{5}
\; \mbox{ and } \; \nabla_{\mu} {\bf e}^{5} = 0.
\end{displaymath}
Finally, it is a simple matter to show that in the basis of five-tensors of
rank $(1,1)$ associated with the $P$-basis (55), the components of the
tensor $\TT$ that describes the active Poincare transformation corresponding
to transformation (3) of Lorentz coordinates, are given by formula (5).

\end{document}